 \documentclass[final,5p,times]{elsarticle}

 \usepackage{graphics}

\usepackage{amssymb}

\journal{NIM A  RICAP-2013}

\begin{document}

\begin{frontmatter}



\title{ Estimation of the TeV gamma-ray duty cycle of Mrk 421 with the Milagro observatory }


\author[ia]{B. Patricelli}
\author[ia]{M.M. Gonz\'alez}
\author[ia]{N. Fraija}
\author[if]{A. Marinelli}
\author{for the Milagro collaboration}

\address[ia]{Instituto de Astronom\'ia, UNAM, M\'exico D.F., 04510, M\'exico}

\address[if]{Instituto de Fisica, UNAM, M\'exico D.F., 04510, M\'exico}

\begin{abstract}
Markarian 421 (Mrk 421) is one the brightest and closest (z=0.031) blazars known (de Vaucouleurs et al 1991 \cite{1991rc3..book.....D}). It is also one of the fastest varying TeV $\gamma$-ray sources, with a flaring activity on time scales as short as tens of minutes. The activity of Mrk 421 at different frequencies may reflect the radiation mechanisms involved. Tluczykont et al. (2007) \cite{2007JPhCS..60..318T} estimated the TeV activity of Mrk 421 through calculating the fraction of time spent in flaring states at TeV energies (TeV duty cycle) by using data from several imaging atmospheric Cherenkov telescopes (IACTs). Since IACT observations are biased towards high flux states they overestimated the TeV duty cycle of Mrk 421. 
Here we propose an alternative approach to calculate the TeV duty cycle of Mrk 421 that takes advantage of the continuous monitoring of the source by the Milagro experiment, a water Cherenkov detector sensitive to primary $\gamma$-rays between 100 GeV and 100 TeV. We present our estimation of the TeV - duty cycle and study its robustness.
\end{abstract}


\end{frontmatter}


\section{Introduction}
\label{intro}
Blazars form the subclass of active galactic nuclei (AGN) that are most commonly detected at very high energies (VHE, E $>$ 100 GeV, Horns 2008 \cite{2008RvMA...20..167H}). They show a strong flux variability, at almost all frequencies of the spectrum, on different time scales, from minutes (see, e.g., Aharonian et al. 2007 \cite{2007ApJ...664L..71A}) to months (see, e.g., von Montigny et al. 1995 \cite{1995ApJ...440..525V}). This large spread in time variability makes it difficult to quantify important  parameters as the duty cycle ($DC$). The $DC$ is defined as the fraction of time spent in a high (ÒflaringÓ) state:
\begin{equation}\label{eq:DC}
DC=\frac{\sum_i t_i}{\sum_i t_i+T_{\rm baseline}}=\frac{T_{\rm flare}}{T_{\rm flare}+T_{\rm baseline}},
\end{equation}
where $t_i$ is the time spent by the source in a $i$ flaring state, $T_{\rm flare}$ is the total time spent by the source in all flaring states ($T_{\rm flare}=\sum_i t_i$) and $T_{\rm baseline}$ is the total time that the source is in the baseline flux state. The baseline flux may be stable and constant with time, although it may present intrinsic variations. In the former case, a flaring state is as any state with flux higher than the baseline flux. In the latter case, a flaring state must be defined taking into account the assumed or measured intrinsic variations of the baseline flux. Thus, a flaring state is defined by a threshold flux and a given energy range, both chosen differently in the literature (see e.g. Krawczynski et al 2004 \cite{2004ApJ...601..151K}, Tluczykont et al. (2007) \cite{2007JPhCS..60..318T} and Wagner 2008 \cite{2008MNRAS.385..119W}).  The identification of a baseline level is also needed to identify the blazar flaring level: without a proper baseline level, only an upper limit of the flaring flux can be determined \cite{2011ICRC....8..147W} . 

Mrk 421 is one of the brightest blazars known and one of the fastest varying $\gamma$-ray sources (Gaidos et al. 1996 \cite{1996Natur.383..319G}). It was the first BL Lac object detected at energies above 100 MeV by EGRET in 1991 (Lin et al. 1992 \cite{1992ApJ...401L..61L}) and the first extragalactic source to be discovered as a TeV emitter by Whipple (Punch et al. 1992 \cite{1992Natur.358..477P}).  

Tluczykont et al. (2007) \cite{2007JPhCS..60..318T}  estimated the TeV duty cycle of Mrk 421. They used data from different IACTs (HEGRA, HESS, MAGIC, CAT, Whipple and VERITAS) from 1992 to 2009. They combined the light curves from these different observatories converting the measured integral flux to flux values in units of the Crab Nebula flux and normalizing to a common energy threshold of 1 TeV and obtained a distribution of flux states for Mrk 421. Finally, they estimated the TeV duty cycle as the ratio between the time that the source spent in a flaring state and the total observation time of the telescopes. They performed the calculation for different flare flux thresholds. For a flare flux threshold of 1 Crab, they found a TeV $DC$ of $\sim$ 40 \%. This value may overestimate the true TeV $DC$ since IACT observations are biased towards high flux states due to their external and self triggering on high states (Tluczykont et al. 2007 \cite{2007JPhCS..60..318T}). In this paper we present a different approach with respect to Tluczykont et al. (2007)  \cite{2007JPhCS..60..318T}  to calculate the TeV $DC$ of Mrk 421 for a flare flux threshold of 1 Crab. This approach takes advantage of the continuous and unbiased long term monitoring by the Milagro detector.

\section{The Milagro detector}
\label{milagro}
Milagro (Atkins et al. 2004 \cite{2004ApJ...608..680A}) was a large water-Cherenkov detector located in the Jemez Mountains near Los Alamos, New Mexico, USA at an elevation of 2630 m above sea level. It was sensitive to extensive air showers resulting from primary gamma rays at energies between 100 GeV and 100 TeV (Abdo et al. 2008a,b \cite{2008ApJ...688.1078A,2008PhRvL.101v1101A}). It had a 2 sr field of view and a 90 \% duty cycle that allowed continuous monitoring of the entire overhead sky. It operated from 2000 to 2008. It was composed of a central 80 m $\times$ 60 m $\times$ 8 m water reservoir instrumented with 723 photomultiplier tubes (PMTs) arranged in two layers. The top ``air-shower'' layer (under 1.4m of purified water) consisted of 450 PMTs, while the bottom ``muon'' layer had 273 PMTs located 6m below the surface. The air-shower layer was used to reconstruct the direction of the air shower by measuring the relative arrival times of the shower particles across the array. The muon layer was used to discriminate between gamma-ray induced and hadron-induced air showers. In 2004, a sparse 200 m x 200 m array of 175 ``outriggers'' was added around the central reservoir. This array increased the area of the detector and improved the gamma/hadron separation. The instrument reached its final configuration (physical configuration, analysis procedures and calibration) in 2005 September.

\section{Estimation of the TeV duty cycle of Mrk 421}
\label{DC}
We analysed data collected by Milagro from September 21, 2005 to March 15, 2008. During this period Mrk 421 was detected with a statistical significance of 7.1 standard deviations at a median energy of 1.7 TeV (Abdo et al., 2013 \cite{MilagroMrk421}). From the study of the light curve we found (Abdo et al. 2013 \cite{MilagroMrk421}) that the Mrk 421 flux is consistent with being constant along the whole 3-year observation period, with an average value above 1 TeV of $\bar f$= ($2.05 \,\pm 0.30$) $\times 10^{-11}\,\rm{cm^{-2}\,s^{-1}}$ ($\chi^2$=134 for 122 degrees of freedom) equivalent to 0.85$\pm$0.13 Crab. This average flux results from time periods where the source is at the baseline state with flux $F_{\rm baseline}$, and periods at any ``flaring'' state $i$, with flux $f_{{\rm flare},i}$. Thus,
\begin{equation}\label{eq:fluence}
\bar{f} \times T_{\rm Milagro}= F_{\rm baseline}\times T_{\rm baseline}+\cal{F_{\rm flare}},
\end{equation}

where $T_{\rm Milagro}$ is the total monitoring period of Milagro given by $T_{\rm baseline}+T_{\rm flare}$  and $\cal{F_{\rm flare}}$ is the total fluence of all high states given by $\sum_i f_{{\rm flare},i}\, t_i$.  

The knowledge of $\bar f$ alone does not allow to estimate the TeV $DC$, as the same value of $\cal{F_{\rm flare}}$ could be obtained by considering many long-duration low-flux flares or a few short-duration high-flux flares, leading to different $DC$ values. Therefore, a distribution of flux flaring states of Mrk 421 is needed. We used the distribution of flux states above 1 TeV reported by Tluczykont et al. (2007,2010) \cite{2007JPhCS..60..318T,2010A&A...524A..48T}. Tluczykont et al. 2010 \cite{2010A&A...524A..48T} found that the distribution above 0.25 Crab can be fit by an exponential function; a better fit of the whole distribution was obtained with a function $f(x)$\footnote{The variable x represents the flux of Mrk 421 above 1 TeV in Crab unit.} which is the sum of a Gaussian component $f_{\rm G}(x)$, describing the baseline flux state plus a log-normal function $f_{\rm Ln}(x)$, describing flaring states (Tluczykont et al. 2010 \cite{2010A&A...524A..48T}):

\begin{equation}\label{eq:fdd}
f(x)=f_{\rm G}(x)+f_{\rm Ln}(x),
\end{equation}
with
\begin{equation}
f_{\rm G}(x)=\frac{N_{\rm G}}{\sigma_{\rm G}\,\sqrt{2 \pi}}\,\exp\left[-\frac{1}{2}\left(\frac{x-\mu_{\rm G}}{\sigma_{\rm G}}\right)^2 \right]
\end{equation}
and
\begin{equation}\label{eq:fln}
f_{\rm Ln}(x)=\frac{N_{\rm Ln}}{x\,\sigma_{\rm Ln}\,\sqrt{2 \pi}} \exp\left[-\frac{({\rm log}(x)-\mu_{\rm Ln})^2}{2 \sigma_{\rm Ln}^2} \right].
\end{equation}

The mean of the Gaussian component, $\mu_{\rm G} \sim$0.33 Crab, represents an upper limit on the value of $F_{\rm baseline}$ (Tluczykont et al. 2010 \cite{2010A&A...524A..48T}). In fact, lower fluxes may be missing in the distribution due to the fact that the detectors used may not be sensitive enough to detect them for short observation periods.

The function $f(x)$ can be used to calculate the average flare flux of Mrk 421, $< f_{\rm flare} >$:
\begin{equation}\label{eq:ffm}
<f_{\rm flare}>=\frac{\int_{1 \, \rm{Crab}}^{F_{\rm lim}} x\,f(x)\,dx}{\int_{1 \, \rm{Crab}}^{F_{\rm lim}} f(x)\,dx}
\end{equation}

where $F_{\rm lim}$ is the maximum flux considered in the distribution, i.e. $F_{\rm lim}$=10 Crab \cite{2010A&A...524A..48T} (here we are considering a flare flux threshold of 1 Crab).  Then, we have $<f_{\rm flare}>$= 2.64 Crab.

${\cal F_{\rm flare}}$ can be written in terms of $<f_{\rm flare}>$ as:

\begin{equation}\label{eq:fluencehigh}
{\cal F_{\rm flare}}=<f_{\rm flare}>\times T_{\rm flare}.
 \end{equation}
By inserting Eq. \ref{eq:fluencehigh} in Eq. \ref{eq:fluence} we obtain
\begin{equation}\label{eq:tflare}
T_{\rm flare}=\frac{\left( \bar{f}-F_{\rm baseline}\right) T_{\rm Milagro}}{<f_{\rm flare}>-F_{\rm baseline}}
\end{equation}
Then, Eq. \ref{eq:DC} becomes,

\begin{equation}\label{eq:DC2}
DC=\frac{\left( \bar{f}-F_{\rm baseline}\right)}{<f_{\rm flare}>-F_{\rm baseline}}.
\end{equation}
From Eq. \ref{eq:DC2} it is clear that the TeV $DC$ depends on three quantities: 1) the average flux of Mrk 421 ($\bar f$) which has a unique value of 0.85$\pm$0.13 Crab as determined by Milagro observations; 2) the value of the baseline flux ($F_{\rm baseline}$), known to be in the range between 0 and the maximum value of 0.33 Crab and; 3) the average flare flux $< f_{\rm flare}>$ that mainly depends on the flaring state distribution (i.e., on $f(x)$). In particular, as we considered flares with a flux greater than 1 Crab, the only component of $f(x)$ involved in the TeV $DC$ calculation is $f_{\rm Ln}(x)$, with the parameters $\sigma_{\rm Ln}$ and $\mu_{\rm Ln}$ (see Eq. \ref{eq:ffm}). 

\begin{figure}
\includegraphics[width=0.5\textwidth]{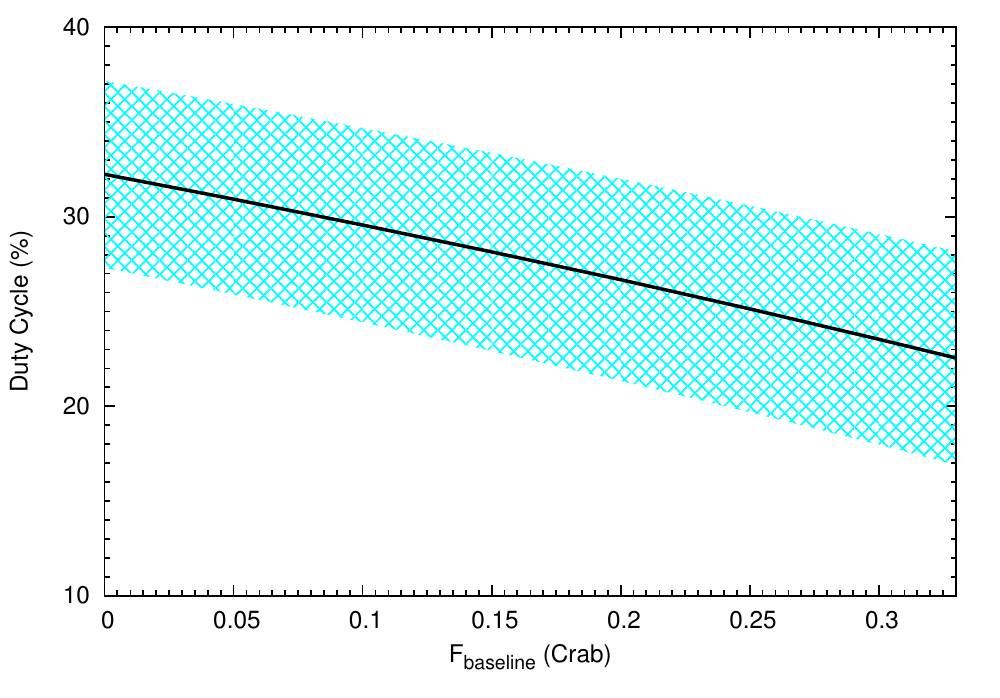}
\caption{Duty cycle calculated by considering as flaring states all those having a flux above 1 TeV greater than 1 Crab. The shadowed blue area represents the error associated to DC, obtained by taking into account the uncertainty on $\bar f$.}\label{fig:1}
\end{figure}

We calculated the TeV $DC$ (see Fig. \ref{fig:1}) for values of $F_{\rm baseline}$ from 0 to the upper limit of 0.33 Crab and the uncertainty due to the error associated to $\bar f$, $\Delta \bar f$. The errors given by the uncertainties on the parameters of $f(x)$ are discussed in Sec. \ref{sec:sigma} and \ref{sec:mu}. In Sec. \ref{sec:exp} we also show the calculation of the TeV $DC$ using, instead of $f(x)$, the exponential function given by Tluczykont et  al. 2010 \cite{2010A&A...524A..48T}.

From Fig. \ref{fig:1} it can be seen that the TeV $DC$ ranges from $22.6^{+ 5.6}_{-5.7}$ \% ($F_{\rm baseline}$=0.33 Crab) to $32.2^{+5.0}_{-4.9}$ \% ($F_{\rm baseline}$=0 Crab). These values are lower than, but marginally consistent within the error with the 40 \% value obtained by Tluczykont et al. (2007) \cite{2007JPhCS..60..318T}. This result is not surprising since, as already explained in Sec. \ref{intro}, the calculation by Tluczykont et al. (2007) \cite{2007JPhCS..60..318T} is affected by an observational bias to continue observations of the source in high states, that leads to an overestimate of the TeV $DC$.

\subsection{Uncertainty in the $\sigma_{\rm Ln}$ parameter}\label{sec:sigma}
\begin{figure}[h!]
\includegraphics[width=0.5\textwidth]{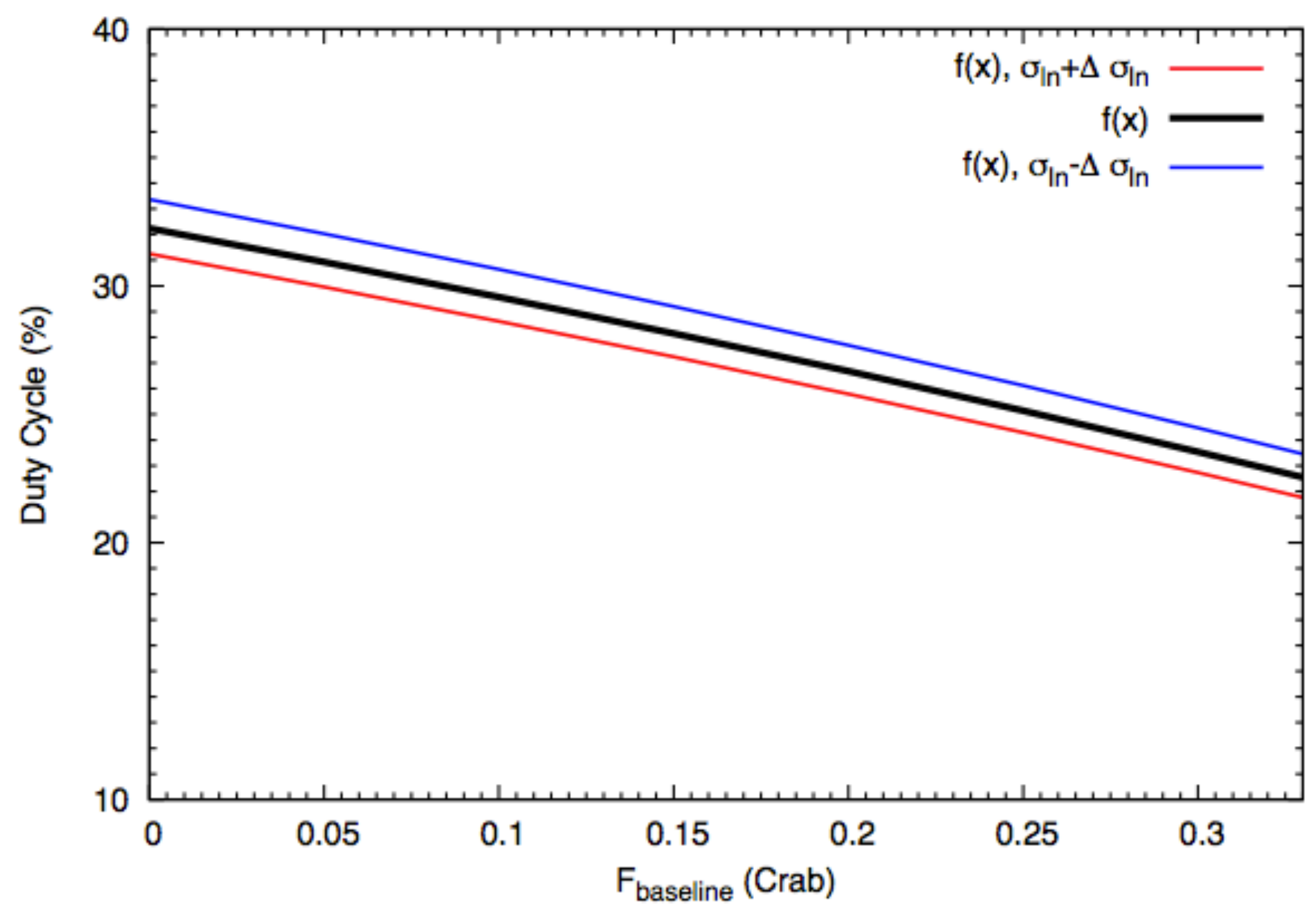}
\caption{Duty cycle calculated by considering as flaring states all those having a flux above 1 TeV greater than 1 Crab. The black line corresponds to the calculation done by assuming the best fit values for the parameters of $f(x)$ (see text); the red and the blue lines correspond to the calculation done by assuming $\sigma_{\rm Ln}$ + $\Delta \sigma_{\rm Ln}$ and $\sigma_{\rm Ln}$ - $\Delta \sigma_{\rm Ln}$ respectively, with $\Delta \sigma_{\rm Ln}$ the error associated to $\sigma_{\rm Ln}$ (Tluczykont et al. 2010 \cite{2010A&A...524A..48T}).}\label{fig:2}
\end{figure}

We calculated the TeV $DC$ by taking into account the uncertainty on the value of $\sigma_{\rm Ln}$ as reported in Tluczykont et al. (2010) \cite{2010A&A...524A..48T}; the results are shown in Fig. \ref{fig:2}. It can be seen that in this case the TeV $DC$ ranges from $22.6^{+0.9}_{-0.7}$ \% ($F_{\rm baseline}$=0.33 Crab) to $32.2^{+1.2}_{-1.0}$ \% ($F_{\rm baseline}$=0.0 Crab). The maximum error on $DC$ associated to the uncertainty on $\sigma_{\rm Ln}$ is of the order of 4\% and it is lower than the one due to $\Delta \bar f$.

\subsection{Uncertainty in the $\mu_{\rm Ln}$ parameter}\label{sec:mu}
\begin{figure}[h!]
\includegraphics[width=0.5\textwidth]{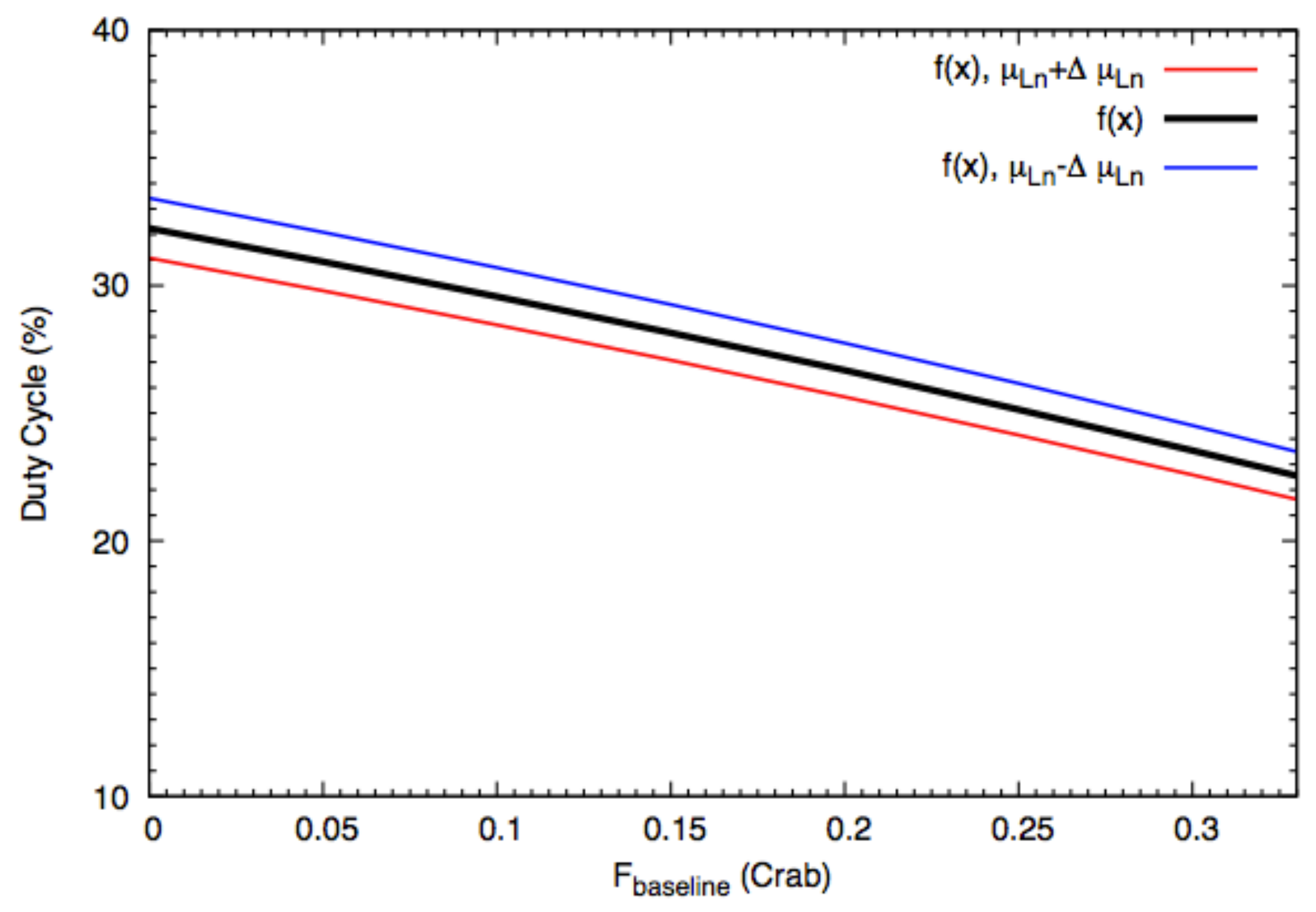}
\caption{Duty cycle calculated by considering as flaring states all those having a flux above 1 TeV greater than 1 Crab. The black line corresponds to the calculation done by assuming the best fit values for the parameters of $f(x)$ (see text); the red and the blue lines correspond to the calculation done by assuming $\mu_{\rm Ln}$ + $\Delta \mu_{\rm Ln}$ and $\mu_{\rm Ln}$ - $\Delta \mu_{\rm Ln}$ respectively, with $\Delta \mu_{\rm Ln}$ the error associated to $\mu_{\rm Ln}$ (Tluczykont et al. 2010 \cite{2010A&A...524A..48T}).}\label{fig:3}
\end{figure}

We calculated the TeV $DC$ by taking into account the uncertainty on the value of $\mu_{\rm Ln}$, as reported in Tluczykont et al. (2010) \cite{2010A&A...524A..48T}; the results are shown in Fig. \ref{fig:3}. It can be seen that in this case the TeV $DC$ ranges from $22.6^{+1.0}_{-0.9}$ \% ($F_{\rm baseline}$=0.33 Crab) to $32.2^{+1.2}_{-1.1}$ \% ($F_{\rm baseline}$=0.0 Crab). The maximum error on $DC$ associated to the uncertainty on $\mu_{\rm Ln}$ is of the order of 4\% and it is lower than the one due to $\Delta \bar f$.

\subsection{Exponential function}\label{sec:exp}
We calculated the TeV $DC$ also by using, instead of $f(x)$, the exponential function in Tluczykont et al (2010) \cite{2010A&A...524A..48T}. In this case we found that $DC$ ranges from 33.4 \% ($F_{\rm baseline}$=0,33 Crab) to 45.1 \% ($F_{\rm baseline}$=0.0 Crab). The higher values of the TeV $DC$ with respect to the ones obtained with $f(x)$ are a consequence of the fact that the exponential function underestimates the number of flares with flux above a few Crab (see Fig. 3 Tluczykont et al. 2010 \cite{2010A&A...524A..48T}). Therefore, the estimated $< f_{\rm flare} >$ is lower and the source should have been in a flaring state for a greater
 time in order to have a total fluence equal to the one observed by Milagro ( $\bar f \times T_{\rm Milagro}$, see also Eq. \ref{eq:tflare}). Therefore the use of the exponential function leads to an overestimate of the TeV $DC$.

\begin{figure}[h!]
\includegraphics[width=0.5\textwidth]{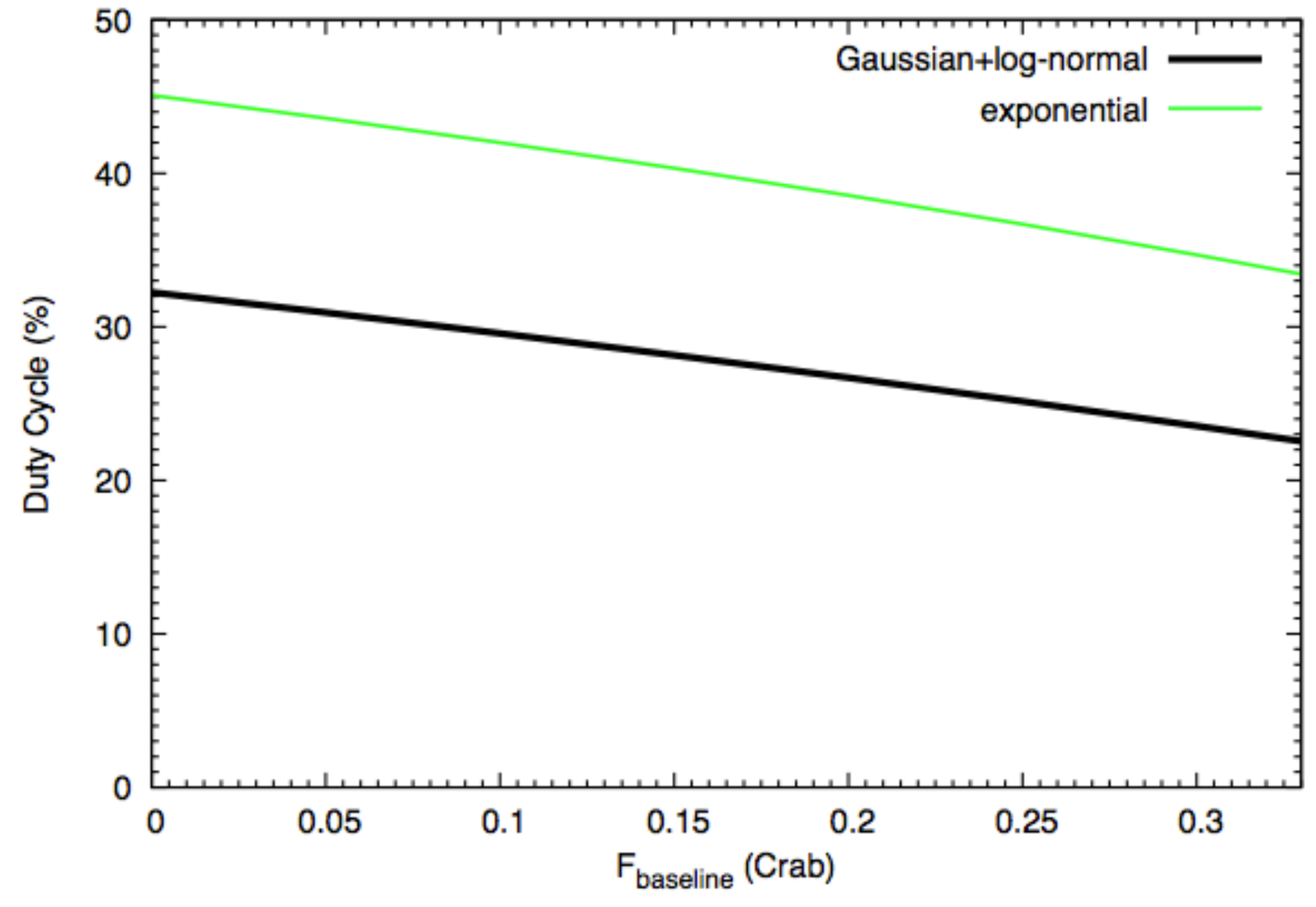}
\caption{Duty cycle calculated by considering as flaring states all those having a flux above 1 TeV greater than 1 Crab. The black line corresponds to the calculation done by using the log-normal plus the Gaussian function (see text); the green line correspond to the calculation done by using the exponential function (Tluczykont et al. 2010 \cite{2010A&A...524A..48T}).}\label{fig:4}
\end{figure}

\section{Conclusions}
We have presented a new approach to estimate the TeV $DC$ of Mrk 421, that takes advantage of the continuous monitoring of the source with the Milagro experiment. We have considered the activity of the source above 1 Crab at TeV energies and we found that, depending on the assumed value for the baseline flux of Mrk 421, the TeV $DC$ ranges from $22.6^{+ 5.6}_{-5.7}$ \% to $32.2^{+5.0}_{-4.9}$ \%. These values are lower than but consistent, within the errors, with the value found by Tluczykont et al. 2007 \cite{2007JPhCS..60..318T}. We also tested the robustness of the calculation, taking into accout the uncertainties in the parameters of the log-normal function describing the distribution of flux states of Mrk 421 (Tluczykont et al. 2010 \cite{2010A&A...524A..48T}). We found that the maximum error on the $DC$ due to these uncertainties is 4 \%. This error is much lower than the one associated to the uncertainty on the average flux observed by Milagro. Finally, we have shown that the use an exponential function instead of the log-normal function leads to an overestimation of the TeV $DC$. 

The value of 1 Crab chosen as flare flux threshold represents an overestimate of the minimum flux required to define a flaring state: in fact, Tluczykont et al. 2010 \cite{2010A&A...524A..48T} have pointed out that above a few tenths of Crab the distribution of flux states presents the typical behaviour of ``high'' states. The estimation of the TeV $DC$ for more realistic assumptions on the threshold flare flux will be presented elsewhere, together with a comparison of the TeV $DC$ with the X-ray $DC$.

\vspace*{0.5cm}
\footnotesize{{\bf Acknowledgment:}{We gratefully acknowledge Scott Delay and Michael Schneider for their dedicated efforts in the construction and maintenance of the Milagro experiment. This work has been supported by the Consejo Nacional de Ciencia y Tecnolog\'ia (under grant Conacyt 105033), Universidad Nacional Aut\'onoma de M\'exico (under grants PAPIIT IN105211 and IN108713) and DGAPA-UNAM.}}





\bibliographystyle{elsarticle-num}



\end{document}